\documentclass[conference]{IEEEtran}

\UseRawInputEncoding
\usepackage{mathtools}
\usepackage{multicol}
\usepackage{amsmath,amssymb,cite,multirow,subfigure,amstext}
\usepackage{graphicx}
\usepackage{subfigure}
\usepackage{url}
\usepackage{multirow}
\usepackage[linesnumbered,ruled, vlined]{algorithm2e}
\usepackage{algorithmic}
\usepackage{makecell}
\usepackage{cite}
\usepackage{bm}

\usepackage{booktabs}
\usepackage{threeparttable}

\usepackage{fancyhdr}
\usepackage[marginal]{footmisc}

\begin{document}
%
\title{Accelerate Three-Dimensional Generative Adversarial Networks Using Fast Algorithm }


\author{\IEEEauthorblockN{Ziqi Su$^1$, Wendong Mao$^1$, Zhongfeng Wang$^1$, Jun Lin$^1$, Wenqiang Wang$^2$, and Haitao Sun$^2$}\\
\IEEEauthorblockA{$^1$School of Electronic Science and Engineering, Nanjing University, P. R. China\\
$^2$ SenseTime Research\\
Email: \{zqsu,wdmao\}@smail.nju.edu.cn, \{zfwang,jlin\}@nju.edu.cn, \{wangwenqiang, sunhaitao\}@powertensors.ai}}
\maketitle
\begin{abstract}
Three-dimensional generative adversarial networks (3D-GAN) have attracted widespread attention in three-dimension (3D) visual tasks. 3D deconvolution (DeConv), as an important computation of 3D-GAN, significantly increases computational complexity compared with 2D DeConv. 3D DeConv has become a bottleneck for the acceleration of 3D-GAN. Previous accelerators suffer from several problems, such as large memory requirements and resource underutilization. To handle the above issues, a fast algorithm for 3D DeConv (F3DC) is proposed in this paper. F3DC applies a fast algorithm to reduce the number of multiplications and achieves a significant algorithmic strength reduction. Besides, F3DC removes the extra memory requirement for overlapped partial sums and avoids computational imbalance to fully utilize resources. Moreover, we design an F3DC-based hardware architecture, which consists of four fast processing units (FPUs). Each FPU includes a pre-process module, a EWMM module and a post-process module for F3DC transformation. By implementing our design on the Xilinx VC709 platform for 3D-GAN, we achieve a throughput up to 1700 GOPS and 4$\times$ computational efficiency improvement compared with prior works.
\end{abstract}

\section{Introduction}\label{sec:intro}
With the development of deep learning,  three-dimension (3D) deep neural
networks have been widely used in many visual fields. Among them, Three-Dimension Generative Adversarial Networks (3D-GAN) have been used for various visual tasks, such as 3D object recognition and reconstruction~\cite{DBLP:journals/corr/0001ZXFT16,8265295}, 3D model generation~\cite{DBLP:journals/corr/0001ZXFT16}, medical image analysis~\cite{DBLP:journals/corr/abs-2003-13653,cicek_3d_2016,7950542}, human action recognition~\cite{6165309}, and so on. 3D-GAN usually contain operations like 3D convolution, 3D deconvolution (DeConv), which is also called transposed convolution.
\footnote{This work was supported in part by the National Natural Science Foundation of China under Grant 62174084, 62104097 and in part by the High-Level Personnel Project of Jiangsu Province under Grant JSSCBS20210034, the Key Research Plan of Jiangsu Province of China under Grant BE2019003-4. (Corresponding author: Zhongfeng Wang; Jun Lin.)}
Many works~\cite{shen2018acm,fangchao} investigated the acceleration of 3D convolution for 3D Convolutional Neural Networks (CNNs). For example, Winograd algorithm is adopted in~\cite{shen2018acm} to propose a template-based methodology for 2D and 3D CNNs and designed corresponding accelerators.~\cite{fangchao} introduced a fast FFT-based algorithm and F3D-based hardware architecture for 3D CNNs, which significantly reduced the computational complexity of 3D convolution operations. Even though the basic computation schemes of 3D convolution and 3D DeConv are similar, 3D DeConv needs to insert zeros into the original input feature maps before performing normal convolution. However, directly applying the methods of 3D convolution for DeConv leads to severe hardware underutilization.

Besides, some works~\cite{di_exploring,wang_towards,scale-out_shen} proposed efficient solutions to accelerate 2D DeConv. However, 3D DeConv has higher computational complexity compared with 2D DeConv, which makes it hard to accelerate 3D DeConv by 2D methods. For instance,~\cite{di_exploring} presented a novel Wino-transCONV dataflow to accelerate 2D DeConv, but 3D DeConv is more complicated than 2D DeConv, which makes it hard to employ DeConv dataflow.~\cite{wang_towards} applied input oriented mapping (IOM) method to accelerate 3D DeConv. However, the generated overlapped partial sums will increase along with the size of kernels by using the IOM method, which leads to increased computing resources and complex dataflow.~\cite{scale-out_shen} accelerated 3D DeConv by taking advantage of the fixed sparsity pattern of intermediate data tiles, but resulted in unbalanced computations. 

To tackle these problems, we propose a fast algorithm named F3DC to accelerate 3D DeConv. Specifically, the contributions of this paper are concluded as follows.
\begin{itemize}
\item By investigating the mathematical formation of 3D DeConv, we propose an efficient computation method, namely F3DC, for 3D DeConv based on a fast transformation algorithm~\cite{MAO}, which greatly reduces the computational complexity and simplifies the computation flow.
\item Based on the F3DC, we develop an efficient architecture to implement 3D DeConv layers. Fast processing unit and fast processing array are designed to implement F3DC transformation and improve parallelism, respectively.

\item The proposed design is implemented on the Xilinx ZC709 platform and achieves a computational throughput of 1700 GOPS and up to $4\times$ improvement on computational efficiency compared with prior works.
\end{itemize}

\section{Background}\label{sec:bg}
\begin{figure}[hbt]
\centering
    \includegraphics[width=3.5in]{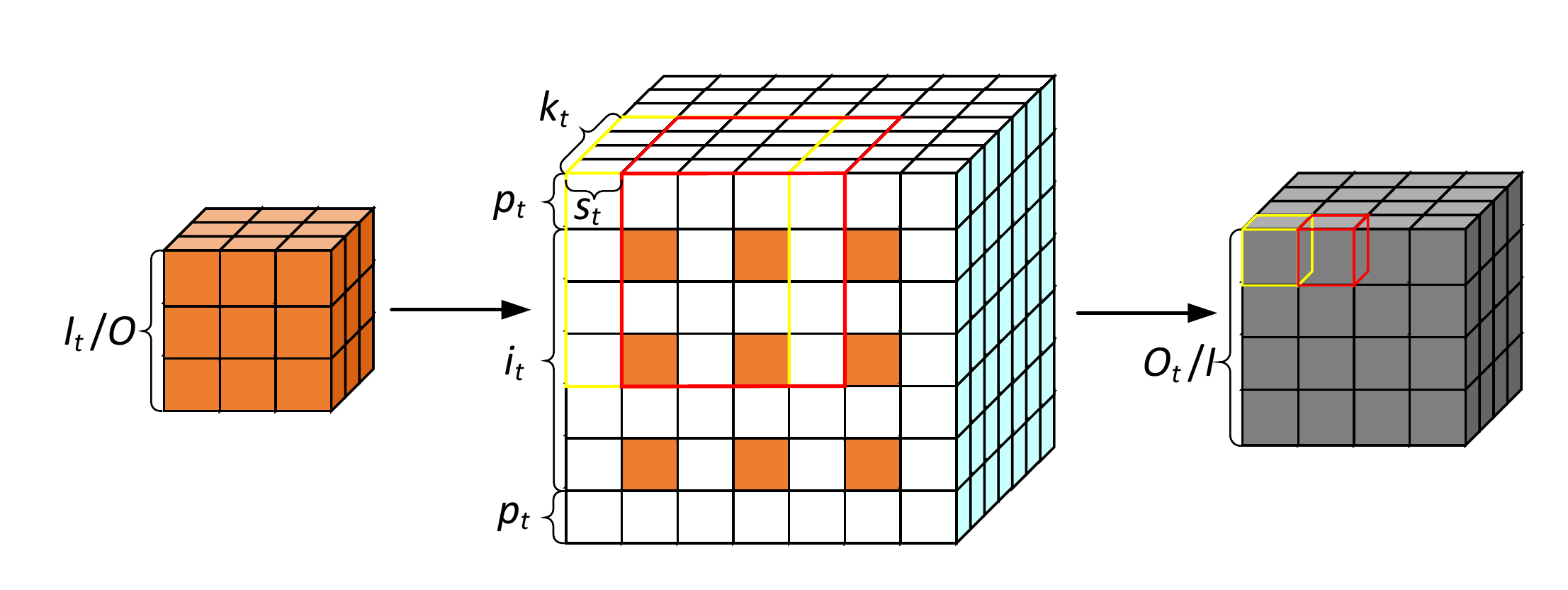}
    \caption{Illustration of 3D deconvolution.}
\label{fig:3DTransConv}
\end{figure}

Compared with 3D convolution, the 3D DeConv needs to insert zeros before performing normal convolution. Fig.~\ref{fig:3DTransConv} shows the process of 3D DeConv.

DeConv is used to expand the input characteristics, which is different from convolution. For the convolution process, the relationship between input feature map and output feature map is represented by the following Eq.~\eqref{BC}:
\begin{small}
\begin{equation}\label{BC}
o=\left\lfloor\frac{i+2 p-k}{s}\right\rfloor+1,
\end{equation}
\end{small}
where $i$ and $o$ represent the size of input and output feature map, respectively, and $k$ means the kernel size of the convolution. $p$ and $s$ denote the padding and stride.

As illustrated in Fig.~\ref{fig:3DTransConv}, multiple parameters are introduced to describe 3D DeConv. The relationship of parameters are shown by the following Eq.~\eqref{BC2}:
\begin{small}
\begin{equation}\label{BC2}
\begin{aligned}
&i_t=o+(s-1)(o-1),\\
&p_t=\frac{(k-1)+(k-2 p-1)}{2}=k-p-1,\\
&k_t=k, \\
&s_t=1, \\
&O_t=i=\frac{i_t+2 p_t-k_t}{s_t}+1,
\end{aligned}
\end{equation}
\end{small}
where $i_t$ represents the result size of inner zero inserting, $p_t$, $k_t$ and $s_t$ means the padding, kernel and stride for DeConv, respectively. $O_t$ means the size of DeConv output.

The inserted zeros occupy more than seventy-five percent of data in 3D DeConv, which results in a large amount of invalid operations. Avoiding invalid operations is necessary to reduce computational complexity.

Some works~\cite{wang_towards,di_exploring} focused on accelerating transposed convolution. For example, Wang \emph{et al.}~\cite{wang_towards} introduced the IOM method to accelerate 3D DeConv. However, the overlapped partial sums of the IOM lead to increased computing resources and complex dataflow. Di \emph{et al.}~\cite{di_exploring} presented the novel Wino-transCONV dataflow and corresponding hardware architecture design, but the data rearrangement process is more complicated for 3D DeConv. Hence, we propose the F3DC to tackle the inserted zeros and reduce computational complexity.

\section{F3DC Algorithm}\label{sec:ERA}
\subsection{Computational Procedure}\label{ssec:FRU}
Since 3D DeConv has higher computational complexity than 2D DeConv, it is necessary to exploit fast algorithms to reduce the algorithmic complexity. Hence, we design F3DC, a computationally efficient 3D DeConv algorithm based on fast transformation algorithm (FTA)\cite{MAO}. 

FTA is an algorithm to transform 2D DeConv into matrix multiplications by the following Eq.~\eqref{2DTRAN}:
\begin{equation}\label{2DTRAN}
\begin{aligned}
\mathbf{Y}=\mathbf{A}^{\mathrm{T}}\left[\left(\mathbf{H} \cdot \mathbf{g} \cdot \mathbf{H}^{\mathbf{T}}\right) \odot\left(\mathbf{P}^{\mathrm{T}} \cdot \mathbf{d} \cdot \mathbf{P}\right)\right] \mathbf{A},
\end{aligned}
\end{equation}
where $g$ is a $k\times k$ 2D DeConv kernel and $d$ is an $I_r\times I_r$ 2D input tile. However, 3D DeConv has an extra dimension compared with 2D DeConv. The extra dimension leads to 2D methods can not directly exploit the acceleration potential of 3D DeConv. To tackle this problem, F3DC is developed to specifically accelerate 3D DeConv, which avoids zeros insertion and further reduces computational complexity.

The computation process of F3DC is shown in Eq.~\eqref{3order}. Multiply and accumulate (MAC) are transformed to element-wise multiplication (EWMM) by transformation matrix during the process. The process reduces the number of multiplications and results in lower computational complexity.
\begin{equation}\label{3order}
\begin{aligned}
\mathbf{Y}=\big\{\mathbf{A}^{\mathbf{T}}\big\{\big[\big(\mathbf{H} \cdot \mathbf{g} \cdot \mathbf{H}^{\mathbf{T}}\big)^{\mathbf{R}}\cdot \mathbf{H}^{\mathbf{T}}\big] \\
\odot \big[\big(\mathbf{P}^{\mathbf{T}} \cdot \mathbf{d} \cdot \mathbf{P}\big)^{R} \cdot \mathbf{P} \big] \big\} \mathbf{A}\big\}^\mathbf{CR} \mathbf{A}.
\end{aligned}    
\end{equation}
In Eq.~\eqref{3order} $\odot$ represents EWMM. $g$ is a $k\times k\times k$ 3D DeConv kernel and $d$ is an $I_r\times I_r\times I_r$ 3D input tile. $\mathbf{H}$ denotes an $E_r\times k$ matrix to preprocess 3D kernels and $\mathbf{P}^{\mathbf{T}}$ denotes an $E_r\times I_r$ matrix to preprocess input tiles. $\mathbf{A}^{\mathbf{T}}$ represents an $O_r\times E_r$ post-processing matrix to obtain final output tile, where $I_r=\lceil( k+r\times s-1)/s\rceil$, $E_r=k+(r-1)\times s$, $O_r=s\times r$. $r$ means the order of transformation. $R$ and $CR$ denote the clockwise and counterclockwise rotation, respectively.


\begin{figure*}[hbt]
\centering
    \includegraphics[width=7in]{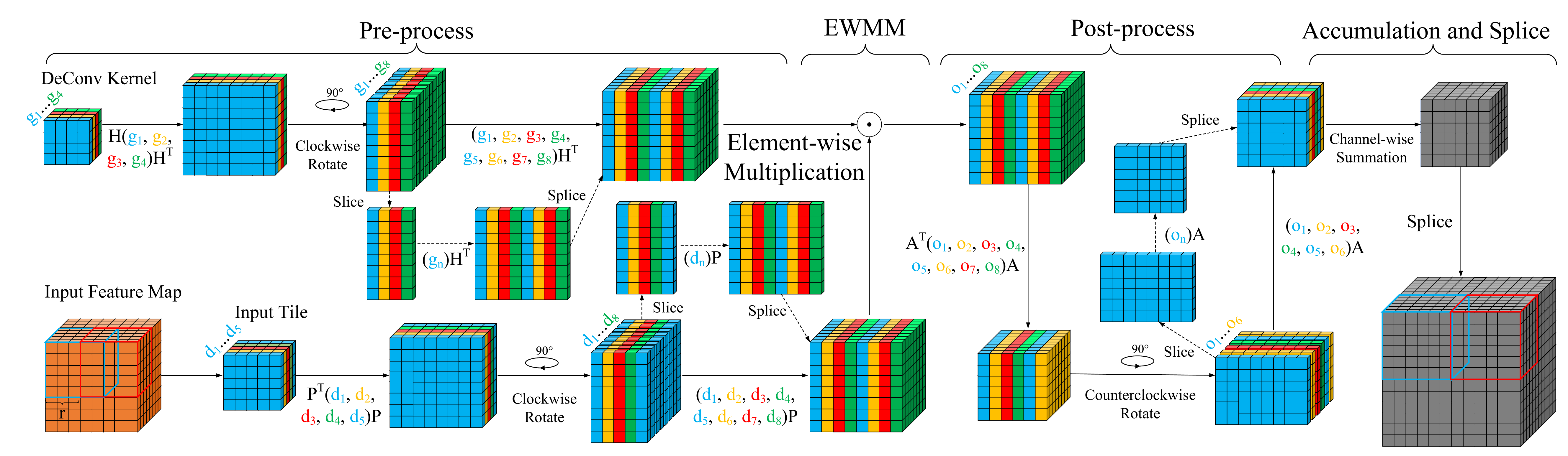}
    \caption{An example of F3DC computational procedure, where the DeConv kernel size is 4 and the stride is 2, $r$ represents the order of F3DC.}
\label{fig:3DLONG}
\end{figure*}

Fig.~\ref{fig:3DLONG} shows the whole process for F3DC. The computation of F3DC is presented in the following four procedures.
\textbf{Pre-process}: the pre-process is shown in Fig.~\ref{fig:3DLONG}. The DeConv kernel and input tiles are sliced for matrix multiplication by $\mathbf{H}$ and $\mathbf{P}^{\mathbf{T}}$ firstly, and then a $90$ degrees clockwise rotation with vertical axis is performed to involve depth dimension before repeating slice and matrix multiplication. By above process, the size of DeConv kernel and input tiles are transformed from $k\times k\times k$ and $I_r\times I_r\times I_r$ to $E_r\times E_r\times E_r$, respectively. For the rotation, because the matrix could not directly conduct multiplication with 3D cubes, the cubes need to be sliced for matrix multiplication, which distinguishes 2D and 3D DeConv computation methods. Besides, 3D DeConv needs to exploit extra dimension for acceleration compared with 2D computation, and as a result, rotation is a necessary procedure for 3D DeConv. 

\begin{footnotesize}
\begin{equation}\label{matrix_5}
\begin{gathered}
\mathbf{P}^{\mathbf{T}}=\left[\begin{array}{ccccc}
1 & 0 & -1 & 0 & 0 \\
0 & 1 & 1 & 0 & 0 \\
0 & -1 & 1 & 0 & 0 \\
0 & -1 & 0 & 1 & 0 \\
0 & 1 & 0 & -1 & 0 \\
0 & 0 & 1 & 1 & 0 \\
0 & 0 & -1 & 1 & 0 \\
0 & 0 & -1 & 0 & 1
\end{array}\right]
\mathbf{H}=\left[\begin{array}{cccc}
0 & 0 & 0 & 1 \\
0 & \frac{1}{2} & 0 & \frac{1}{2} \\
0 & -\frac{1}{2} & 0 & \frac{1}{2} \\
0 & 1 & 0 & 0 \\
0 & 0 & 1 & 0 \\
\frac{1}{2} & 0 & \frac{1}{2} & 0 \\
-\frac{1}{2} & 0 & \frac{1}{2} & 0 \\
1 & 0 & 0 & 0
\end{array}\right] \\
\mathbf{A}^{\mathbf{T}}=\left[\begin{array}{cccccccc}
1 & 1 & 1 & 0 & 0 & 0 & 0 & 0 \\
0 & 0 & 0 & 0 & 1 & 1 & 1 & 0 \\
0 & 1 & -1 & 0 & 0 & 0 & 0 & 0 \\
0 & 0 & 0 & 0 & 0 & 1 & -1 & 0 \\
0 & 1 & 1 & 1 & 0 & 0 & 0 & 0 \\
0 & 0 & 0 & 0 & 0 & 1 & 1 & 1
\end{array}\right]
\end{gathered}
\end{equation}
\end{footnotesize}
\textbf{EWMM}: the two transformed cubes perform EWMM to get the resulting cube for post-processing. EWMM performs $E_r\times E_r\times E_r$ multiplications for each transformation. EWMM is presented in Fig.~\ref{fig:3DLONG}.

\textbf{Post-process}: the results of EWMM are sliced to perform multiplication with post-processing matrix $\mathbf{A}^{\mathbf{T}}$, then a 90 degrees counterclockwise rotation with vertical axis is conducted before repeating slice and matrix multiplication. The opposite direction of rotation is due to the 3D DeConv data arrangement. The result size of post-process is $O_r\times O_r\times O_r$. Post-process is presented in Fig.~\ref{fig:3DLONG}.

\textbf{Accumulation and Splicing}: the result tiles are accumulated channel-wise and spliced for the output feature maps.

The transformation matrix of different kernels could refer to~\cite{MAO}. Here, we use $T_r(o_t^3, k_t^3)$ to denote an r-order F3DC, where $o_t^3$ and $k_t^3$ represent the size of the output and kernel, respectively. Considering $4\times4\times4$ is a common size for 3D DeConv kernel, a usual example of 3-order 3D DeConv $T_3(6^3,4^3)$ transformation matrix is presented in Eq.~\eqref{matrix_5} for clarity.

\subsection{Complexity analysis}\label{ssec:FRU}

Table~\ref{tab:ALG} presents a comparison between F3DC, zero-inserting method (ZIM), and winograd-based method\cite{scale-out_shen} on algorithmic reduction towards 3D DeConv. A winograd-based algorithm is adopted in\cite{scale-out_shen} to compute 3D DeConv on zero-inserted feature maps. The method only removes zeros in the edges leading to insufficiently exploits acceleration potential of fast algorithm, and can not reach the optimal speedup. The arithmetic complexity of F3DC for r-order is denoted as $\mu(T_r[o_t^3, k_t^3])$, it is computed as:
\begin{equation}
\mu\left[T_{r}\left(O_{r}^{3}, k^{3}\right)\right]=\frac{[k+(r-1) \times s]^{3}}{(r \times s)^{3}}.
\end{equation}

As shown in Table~\ref{tab:ALG}, F3DC achieves $27\times$ reduction on multiplications per output compared with ZIM in $k=4$, $s=2$, which significantly improves computational efficiency for 3D DeConv. 
\newcommand{\tabincell}[2]{\begin{tabular}{@{}#1@{}}#2\end{tabular}}
\begin{table}[hbt]
\centering
\caption{algorithmic complexity comparsion$^{*}$}\label{tab:ALG}
\begin{tabular}{|c|c|c|c|c|}
\hline
  &\textbf{\tabincell{c}{$k=3$\\ $s=2$}}  &\textbf{\tabincell{c}{$k=4$ \\$s=2$}}  &\textbf{\tabincell{c}{$k=5$\\ $s=2$}} &\textbf{\tabincell{c}{$k=9$\\ $s=2$}} \\ 
\hline
\textbf{ZIM} &\tabincell{c}{27}  &\tabincell{c}{64}  &\tabincell{c}{125}  &\tabincell{c}{729}\\ 
\hline
\textbf{Winograd-based\cite{scale-out_shen}} &3.375  &8  &15.625  &91.125\\ 
\hline
\textbf{$T_{3}\left(O_{3}^{3}, k^{3}\right)$(Ours)}  &\textbf{1.59} &\textbf{2.37}   &\textbf{3.375} &\textbf{10.17} \\
\hline
\end{tabular}\\\smallskip
\footnotesize{$^*$ The algorithmic complexity indicates the average number of multiplications required to obtain one result.}\\
\end{table}

\section{The Proposed Architecture and Dataflow}\label{sec:TPA}
\subsection{Architecture Overview}
\begin{figure}[hbt]
\centering
    \includegraphics[width=3.5in]{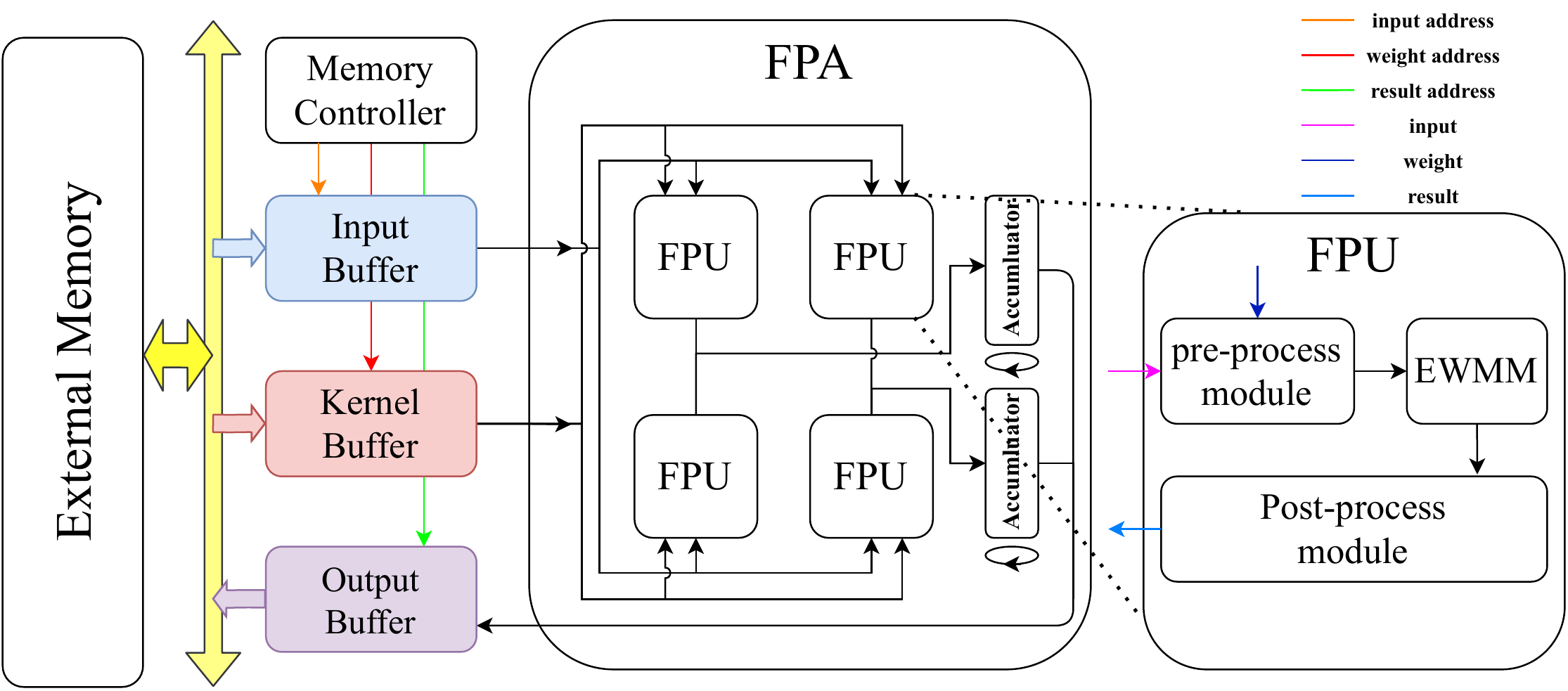}
    \caption{Overview of the proposed hardware architecture.}
\label{fig:hardware_architecture_2}
\end{figure}

As shown in Fig.~\ref{fig:hardware_architecture_2}, the proposed architecture consists of three data buffers (input buffer, kernel buffer, output buffer), fast processing array (FPA), fast processing unit (FPU) and two accumulators. The FPA consists of four FPUs to form $2\times2$ array. Each FPU includes a pre-process module, a EWMM module and a post-process module. The pre-process module executes the transformation of 3D DeConv kernel and input tiles by the matrix of Eq.~\eqref{matrix_5}. Each EWMM module includes $512$ $(8\times8\times8)$ multipliers, and four EWMM modules consume 2048 DSP resources. The post-process module also uses the matrix in Eq.~\eqref{matrix_5} to obtain output tiles before channel-wise summation. The accumulator focuses on the accumulation of output tiles from different channels.
\begin{figure}[hbt]
\centering
    \includegraphics[width=3.5in]{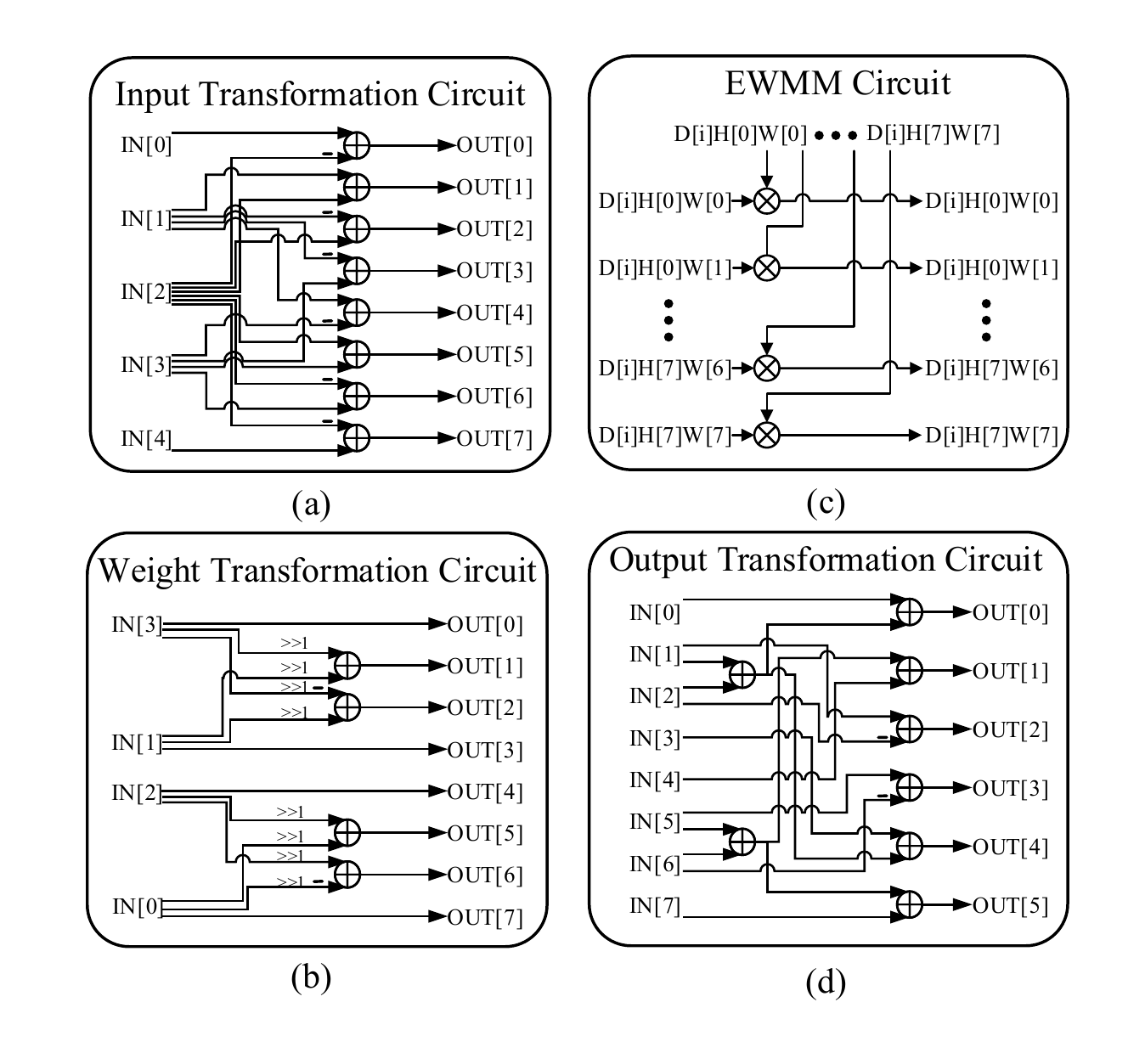}
    \caption{Overall architecture of FPU and the detailed transformation circuit of each module. (a) Input transformation circuit in pre-process module. (b) Weight transformation circuit in pre-process module. (c)EWMM circuit for multiplication. (d) Output transformation circuit in post-process module.}
\label{fig:TC}
\end{figure}
\subsection{FPU Module}
As shown in Fig.~\ref{fig:TC}, an FPU is designed for matrix transformation and EWMM in F3DC by utilizing the simple coefficients in the matrix and the computation for EWMM. The pre-process module implements matrix multiplication responsible for the transformation of kernel and input tiles. Since the number in the transformation matrix mainly consists of ‘1’, ‘-1’, ‘1/2’ and ‘-1/2’. The multiplication of the number can be easily implemented by invert or shift operation, which greatly simplifies the hardware design. The post-process module is similar to the pre-process module in transformation matrix. Besides, the computation procedure for F3DC is suitable for a variety of convolution kernels with little adjustment for the transformation circuits and the number of DSPs of EWMM module. The specified circuit for $T_3(6^3,4^3)$ is shown in Fig.~\ref{fig:TC}.
\subsection{FPA Module}
In order to improve throughput of the architecture, we design the FPA which consists of four FPUs and two accumulators. Four FPUs form a $2\times2$ array to increase parallelism by computing two input channels and two output channels at the same time. For each row of FPUs, they share the same tile from the same input channel, and each column of FPUs computes the same output channel by an accumulator. FPA could execute more computations for one address access and improve data utilization compared with non-parallel design.
\subsection{F3DC Dataflow}
Given that it may not be feasible to load data on chip for all, especially for 3D neural networks. We apply weight stationary\cite{WS} with a loop order of output channel, depth tile, height tile, width tile and input channel. For each input channel loop, the procedure of Fig.~\ref{fig:3DLONG} is computed. Besides, we unroll the loop of the output channel and input channel to increase parallelism.

\section{Experimental Results} \label{sec:results}
\subsection{Experimental Setup}
We choose 3D-GAN\cite{DBLP:journals/corr/0001ZXFT16} model as the benchmark to evaluate our design. The inputs and weights are quantified into 16 bits and 8 bits, respectively. We implement all modules of the architecture with Verilog HDL, and evaluate our architecture on the Xilinx VC709 board with a frequency of 150 MHz. Implementation results are reported by Xilinx Vivado 2020.1.

\subsection{Performance Analysis}

As illustrated in Table~\ref{tab:comp}, we implement a computationally efficient F3DC-based accelerator, which not only executes zero-free operations, but also further reduces the computational complexity. The proposed design achieves 0.83 performance density compared with~\cite{di_exploring}. Even though~\cite{di_exploring} uses a fast algorithm to accelerate DeConv computation, the rearranged filters result in computational imbalance and reduce computation efficiency.~\cite{wang_towards} applies IOM method on 3D DeConv, which means the intermediate results will increase storage overhead. Our method simplifies the dataflow and avoids the storage of intermediate results.~\cite{scale-out_shen} uses Winograd algorithm on the zero-inserted input feature maps to accelerate 3D DeConv, and uses the sparsity of intermediate results to avoid some redundant computations. However,~\cite{scale-out_shen} has complex pre-processing parameters, which increase hardware overhead. The processing parameters of our method are simplified to save hardware resources. In brief, our design can significantly improve computational efficiency and meanwhile reduce hardware complexity.

\begin{table}[hbt]
\centering
\caption{Comparison With Other Works}\label{tab:comp}
\begin{tabular}{|c|c|c|c|c|}
\hline
\textbf{Works}  &\textbf{\tabincell{c}{\cite{fangchao}}}  &\textbf{\tabincell{c}{\cite{di_exploring}}}  &\textbf{\tabincell{c}{~\cite{wang_towards}}} &\textbf{Ours} \\ \hline
\textbf{Platform} &\tabincell{c}{Xilinx\\ VC709}  &\tabincell{c}{Xilinx\\ ZCU102}  &\tabincell{c}{Xilinx\\ VC709}  &\tabincell{c}{Xilinx\\ VC709}\\  \hline
\textbf{Model}  &C3D &3D-GAN  &3D-GAN &3D-GAN \\ \hline
\textbf{Clock(Mhz)}  &200 &200  &200  &150 \\ \hline
\textbf{BRAMs Used}  &1071 &-   &712 &1470 \\ \hline
\textbf{Flip-Flops Used}  &265750 &-  &566182 &212195 \\ \hline
\textbf{LUTs Used}  &257210 &-   &292292 &192342 \\ \hline
\textbf{DSP Used}  &1536 &2520   &2304 &2048 \\ \hline
\textbf{\tabincell{c}{Performance\\(GOPS)}}   &864.1  &482.4  &450$^*$(3600) &1700  \\\hline
\textbf{\tabincell{c}{Performance Density \\(GOPS/DSP)}}  &0.56 &0.19  &0.20 &0.83   \\\hline
\end{tabular}\\\smallskip
\footnotesize{$^*$ Performance is normalized by removing zero-related computations.}\\
\end{table}

\section{Conclusion} \label{sec: conclusion}
In this paper, we first introduce F3DC, a fast algorithm for 3D DeConv, capable of reducing the computational complexity and eliminating invalid operations related to inserted zeros. Furthermore, an efficient hardware architecture is proposed to implement the F3DC-based acceleration of 3D-GAN. Finally, we evaluate our architecture by implementing 3D-GAN model on the Xilinx VC709 platform. The experimental results demonstrate that the proposed architecture can achieve a throughput of 1700 GOPS, which surpasses prior works significantly.

\newpage

\bibliographystyle{IEEEtran}
\bibliography{zqsu}

\end{document}